\documentclass[aps,twocolumn,showpacs,superscriptaddress,amsmath]{revtex4}
\usepackage{graphicx}
\usepackage{amssymb}
\usepackage{bm}

\newcommand{\Eq}[1]{Eq.\,(\ref{#1})}
\newcommand{\be}{\begin{equation}}
\newcommand{\ee}{\end{equation}}

\newcommand{\ra}[1]{\stackrel{\leftarrow}{#1}}
\newcommand{\leftact}{\overset{\rightarrow}}
\newcommand{\rightact}{\overset{\leftarrow}}

\begin{document}

\title{Waiting time distributions of electron transfers through
     quantum dot Aharonov--Bohm interferometers}

\author{Sven Welack}
\email{wesv@ust.hk}
\affiliation{Department of Chemistry, Hong Kong University
  of Science and Technology, Kowloon, Hong Kong}

\author{Shaul Mukamel}
\affiliation{Department of Chemistry, University of California, Irvine, USA}

\author{YiJing Yan}
\email{yyan@ust.hk}
\affiliation{Department of Chemistry, Hong Kong University
  of Science and Technology, Kowloon, Hong Kong}

\date{\today}

\begin{abstract}

We present a statistical readout method for
quantum interferences based on time series analysis of consecutive
single electron transfers through a double quantum dot
Aharonov--Bohm interferometer. Waiting time distributions
qualitatively indicate the presence of
interferences and provide information on orbital-detuning and
coherent interdot-electron transfer.
Interdot transfer induced oscillations are
Aharonov--Bohm  phase sensitive, while
those due to level detuning are phase--independent.
The signature of the quantum interference in the waiting time
distribution is more apparent for weakly coupled electron transfer detectors.

\end{abstract}


\maketitle

%\section{Introduction}

Double-quantum dot (DQD) junctions provide an experimental setup
to study phase coherent transport \cite{Wau95705,Che04176801,Hol0270}
and to realize Aharonov-Bohm (AB) interferometers \cite{Hol01256802,Ihn07111,Sig06036804}.
This is of general interest as they are potential
candidates for qubits. A crucial aspect for their realization
is the noninvasive determination of the presence of
quantum interferences in order to minimize decoherence.

So far theoretical studies on transport properties of double quantum
dot AB interferometers have been focused on average current
\cite{Ape05125302,Mol95125338,Mou05033310,Sim06247207,Tok07113,li00,Kan04117,Kub02245301}
and shot noise \cite{Los001035,Zha06085106,Don08085309,Pen0521452}
properties. Recently time-resolved detection of single electron
transfers in single quantum dots
\cite{Lu03422,Fuj042343,Gus06076605} and DQD in series has become
experimentally feasible \cite{Fuj061634}. Waiting time distributions
of consecutive electron transfers can be obtained from time--series
analysis and provide detailed information on quantum dots
\cite{Wel08195315,Bra08477} and single molecules \cite{Wel081137}.
They were found to be sensitive to interference due to multiple
electron paths in DQD junctions \cite{Wel08195315} and contain more
detailed information than current and noise measurements
\cite{Bra08477}.

In this letter we propose a weakly invasive statistical method based on
waiting time distributions of single external electron transfers
that can determine the presence of quantum interferences,
small detunings of the DQD orbitals, inter-dot transfer coupling,
and Coulomb interaction. These quantities are connected with qualitatively distinguishable oscillations
in the waiting time distribution.
These oscillations are sensitive to the AB phase,
$\phi= \Phi/\Phi_0$, and are suppressed at
$\phi=n \pi$ for an integer $n$ when interdot transfer is present.
Here, $\Phi$ is the magnetic flux
perpendicular to the junction
and $\Phi_0=h/e$ the magnetic flux quanta.
In contrast, the oscillations purely due to energy
detunings of the DQD orbitals are $\phi$--independent.
We show that their detection requires weakly coupled electron detectors
thereby avoiding to inflict fast decoherence on the DQD which qualifies
the proposed method as
a readout scheme for coherently operating qubits.

%The formal result of this letter is a compact notation
To that end, we exploit a master equation in
the many-body Fock space of the DQD,
assuming weak system--reservoir coupling.
%
%\section{Double quantum dot Aharonov Bohm interferometer}
%
For simplicity we consider spinless electrons and each QD dot can
hold only one electron at most. We decompose the total Hamiltonian
of the DQD-AB interferometer junction into
$\tilde{H}_T=\tilde{H}_S+\tilde{H}_R+\tilde{H}_{SR}(\phi)$. The DQD
part (system) reads
\begin{equation}\label{Hs}
\tilde{H}_S=\sum_{s=1,2} \epsilon_{s} c_s^\dag c_{s}
  +U c_1^\dag c_{1} c_2^\dag c_{2}-\Delta (c_1^\dag c_{2} +c_2^{\dag}c_1).
\end{equation}
Here, $\epsilon_{s}$ with $s=1$ or $2$ is the orbital
energy of the specified QD;
$U$ denotes the strength of the Coulomb repulsion between
electrons;
$\Delta$ is the inter-dot electron transfer parameter in
the DQD-AB interferometer.
The Hamiltonian of the electrodes is given by
two independent free electron reservoirs,
$
H_{R}=\sum_{\nu=l,r} \sum_{q} \epsilon_{q \nu} c_{q\nu}^\dag c_{q\nu}.
$
The index $\nu$ denotes the left ($l$) or right ($r$) electrode,
$q$ their intrinsic degrees of freedom.
The electron creation (annihilation) operators
$c^\dag_s$ and $c^\dag_{q\nu}$  ($c_s$ and $c_{q\nu}$)
satisfy the anticommutator relations:
$\{c_k, c_{k^\prime}^\dag\} = \delta_{kk^\prime}$
and $\{c^\dag_k, c^\dag_{k^\prime}\}= \{c_k, c_{k^\prime}\}=0$,
for all $k,k^\prime = s,q\nu$.
The system--reservoirs coupling responsible for
electron transfer between the electrodes and the DQD reads
%is determined by phase dependent single electron transfers
\begin{equation}
\tilde{H}_{SR}(\phi)=\sum_{\nu=l,r} \sum_{sq}
  \left[T_{qs}^{(\nu)}(\phi) c^\dag_s c_{q\nu} + \rm{H.c.} \right].
\end{equation}
The AB phase-dependent transfer parameters
satisfy
$T_{q1}^{(\nu)}(\phi)=T_{q1} e^{i \phi_{\nu}/2}$
and
$T_{q2}^{(\nu)}(\phi)=T_{q2} e^{-i \phi_{\nu}/2}$
for the two parallel dots pierced by a magnetic flux considered here,
with $\phi_l=-\phi_r=\phi$ to
account for the sign change in the
phase between coupling to left and right electrode.
%

%\subsection{Quantum master equation}\label{sec2-master}

We describe the DQD by the reduced density operator
$\rho(t)$ of the system.
Note that
the inter-dot electron transfer $\Delta$ in \Eq{Hs} leads to off-diagonal
elements in the system Hamiltonian $\tilde H_S$ in the orbital basis.
We transform it into eigenbasis, $H_S=O^{-1} \tilde{H}_S O$,
where $O$ consists of the eigenvectors of $\tilde{H}_S$.
The same transformation is applied to the creation
and annihilation operators $\Psi_s=O^{-1} c_s O$ and
$\Psi_s^\dagger=O^{-1} c_s^\dagger O$,
so that $H_{SR}=O^{-1} \tilde{H}_{SR}O$.
The standard perturbation theory leads to
the quantum master equation \cite{Wel08195315}:
\begin{equation}\label{master1}
\dot{ \rho}(t)=-i  {\mathcal L} \rho(t)
 + \sum_{\nu=l,r} (- \Pi_{\nu} +\Sigma^+_{\nu}  + \Sigma^-_{\nu})  \rho(t).
\end{equation}
The system Liouvillian ${\mathcal L}\,\cdot = [H_S,\cdot\,]$
describes the coherent dynamics.
The dissipative superoperator in \Eq{master1}
is separated into the diagonal contribution
$\Pi_{\nu}$ that leaves the number of electrons in the system unchanged,
and the off--diagonal $\Sigma^{+}_{\nu}$ and $\Sigma^{-}_{\nu}$
for the increase and decrease the number of electrons in the DQD,
respectively; see Ref.\ \onlinecite{Wel08195315} for the
derivation.
This separation is necessary in order to keep track
of the trajectories of single electron transfers.
%
%\section{Waiting time distributions}
Consider, for example,
the scenario of detecting an electron
entering the DQD through the left electrode at time $t_0$
and leaving through the right electrode at time $t$.
The waiting-time distribution of consecutive electron transfer
events is then given by the joint-probability \cite{Wel08195315}:
\begin{equation}\label{lr1}
P_{l \rightarrow r}(t, t_0) = \mathrm{tr}_S
   \lbrace \Sigma_r^-  S_{t,t_{0}} \Sigma_l^+ \rho_S(t_0) \rbrace,
\end{equation}
with $S_{t,t_0}= \mathrm{exp} \left[
  \left(-i \mathcal{L}-\Pi_{l}-\Pi_{r} \right) \left(t-t_{0}\right)
 \right]$ being the propagator of the system
in the absence of transfer events at the electrodes
within the waiting time interval.
Quantity (\ref{lr1}) can be obtained from the time--series
of single directionally resolved electron transfers
between the electrodes and the system.
One has to record a sufficiently large number of the
$l \rightarrow r$ events and generate a histogram of
the number of occurrences as function of the
time interval ${t-t_0}$. The histogram has to be
normalized by the total number of considered events.

The aforementioned physically distinct
dissipative components are formally given as \cite{Wel08195315}:
$
\Sigma^+_\nu=\sum_{s}
   \leftact\Psi\,\!_s^\dagger \rightact\Psi\,\!_{\nu s}^{(-)}
 + \rightact\Psi\,\!_s \leftact\Psi\,\!_{\nu s}^{\dagger (-)},
$
$
\Sigma^-_\nu=\sum_{s}
   \rightact\Psi\,\!_s^\dagger \leftact\Psi\,\!_{\nu s}^{(+)}
 + \leftact\Psi\,\!_s  \rightact\Psi\,\!_{\nu s}^{\dagger(+)},
$
and
$ \Pi_{\nu}=\sum_{s}
 \big(\leftact{\Psi}\,\!_{s}^{\dagger}\leftact\Psi\,\!^{(+)}_{\nu s}
 + \rightact\Psi\,\!_{s}^{\dagger}\rightact\Psi\,\!_{\nu s}^{(-)}
 + {\rm H.c.}\big)$.
The involved superoperators are defined
as the left-- or right--actions
($\leftact\Psi\,\cdot\equiv\Psi\,\cdot$
or $\ra\Psi\,\cdot\equiv \cdot\,\Psi$)
of the
associated Hilbert-space $\Psi$--operators.
Besides the annihilation (creation) operators $\Psi_s$ ($\Psi^{\dag}_s$)
we also have to consider their
auxiliaries \cite{Wel06044712,Wel08195315}:
\begin{equation}\label{equ:auxeigen1}
\Psi^{(\pm)}_{\nu s}(t,\phi) = \sum_{s'} \int_{t_0}^t
 \mathrm dt' C^{(\pm)}_{\nu ss'}(t-t';\phi)
 e^{-i \mathcal{L} (t-t')}  \Psi_{s'}.
\end{equation}
Here,
$C^{(+)}_{\nu ss'}(t;\phi)= \sum_{qq'}
 T_{qs}^{(\nu)\ast}(\phi)
 T_{q's'}^{(\nu)}(\phi)
  \langle c_{q\nu}^\dag(t)c_{q'\nu}(0)\rangle_R$
and
$C^{(-)}_{\nu ss'}(t;\phi)= \sum_{qq'}T_{qs}^{(\nu)}(\phi)
 T_{q's'}^{(\nu)\ast}(\phi)
  \langle c_{q\nu}(t)c_{q'\nu}^{\dag}(0)\rangle_R$
are the AB phase--dependent interacting reservoir correlation functions.
 Applying the given phase relations in $T_{q s}^{(\nu)}(\phi)$
and assuming further $T_{q1}=T_{q2}=T_q$ for the AB phase--free
parts lead to the relations:
$
C^{(\pm)}_{\nu 11}(t)=C^{(\pm)}_{\nu 22}(t)=C^{(\pm)}_\nu (t),
$
$
C^{(\pm)}_{\nu 12}(t)=C^{(\pm)}_\nu (t)  e^{i \phi_\nu},
$
and
$
C^{(\pm)}_{\nu 21}(t)=C^{(\pm)}_\nu (t)  e^{-i \phi_\nu}.
$
The auxiliary operators in their non-Markovian form [\Eq{equ:auxeigen1}]
can be numerically evaluated numerically without further approximations
as shown in Ref.\ \onlinecite{Wel08195315} and \onlinecite{Wel06044712}.

To derive analytical results we apply
the Born--Markov approximation, together with
the wide--band limit for the reservoir spectral density.
The latter leads to $C^{(\pm)}_{\nu}(t)=\Gamma \int_0^\infty \mathrm d \epsilon \,
 f_\nu^{(\pm)}(\epsilon) e^{\mp i \epsilon t}$.
Here, $f_\nu^+(\epsilon) = 1-f_\nu^-(\epsilon)
=[e^{(\beta(\epsilon-\mu_\nu)}+1]^{-1}\equiv f(\epsilon-\mu_{\nu})$
is the Fermi distribution function, with
$\beta$ being the inverse temperature and
$\mu_\nu$ the Fermi energy of the electrode $\nu$.
The Born--Markov approximation amounts to
replacing the range of time integration in
\Eq{equ:auxeigen1} with $(-\infty,\infty)$.
As results, the auxiliary annihilation operators
defined in \Eq{equ:auxeigen1} can be evaluated as
\begin{subequations}\label{mark}
\begin{align}
 \Psi^{(\pm)}_{\nu 1}
&= \Gamma f^{\pm}_{\nu}(\mathcal{L})(\Psi_{1}+\Psi_{2} e^{\pm i\phi_{\nu}}),
\\
 \Psi^{(\pm)}_{\nu 2}
&=\Gamma f^{\pm}_{\nu}(\mathcal{L})(\Psi_{1} e^{\mp i\phi_\nu}+\Psi_{2} ),
\end{align}
\end{subequations}
which depend on the AB--phase ($\phi=\phi_{l}=-\phi_r$)
but no long on the time.
The auxiliary creation operators $\Psi^{\dagger(\pm)}_{\nu s}$
are of similar expressions, but with
the replacements of $\mathcal{L} \rightarrow -\mathcal{L}$,
$\phi \rightarrow -\phi$ and $\Psi_s \rightarrow \Psi_s^\dagger$
in \Eq{mark}. Note that since $H_S$ is diagonal in the many--body
Fock space, the action of the superoperator
$f^{\pm}_{\nu}(\mathcal{L})$, which is determined
by the Fermi function and the diagonal system Liouvillian,
can be carried out easily.
All the 16 auxiliary operators, $\Psi^{(\pm)}_{\nu s}$
and $\Psi^{\dagger(\pm)}_{\nu s}$ with $\nu=l,r$ and $s=1,2$,
can now be evaluated [cf.\ \Eq{mark}] in terms of
$4\times$4 matrices in the Fock--space representation.
Consequently, the action of each dissipative tensor
in the second term of \Eq{master1}, which
has been given in terms of the left-- and right--multiplications
of some $\Psi_s$ ($\Psi^{\dag}_s$) and
$\Psi^{(\pm)}_{\nu s}$ ($\Psi^{\dagger(\pm)}_{\nu s}$)
is now determined.
It is worth to mention here
that the approximation scheme explored in \Eq{mark}
leads to an \Eq{master1} in Lindblad form.

%It will be solved numerically
%by Runge-Kutta, with a physical initial condition for $\rho_S(0)$,
%and analytically by using the additional condition
%together with $\mathrm{tr}_S \lbrace \rho_S \rbrace =1$.

We use the following parameter scheme to describe
our calculation results. A bias of $2V$ is applied
symmetrically $\mu_{l/r}=\mu_{\rm eq}\pm V$.
The orbital energies of the DQD are set to be
$\epsilon_{1}=\epsilon_g+\alpha$ and $\epsilon_2=\epsilon_g-\alpha$;
i.e., the orbital energy split (or detuning) is $2\alpha$.
We set the vacuum DQD state $\epsilon_0=0$ as the energy zero,
and $\epsilon_g=1$ the internal energy unit.
In all calculations, $\mu_{\rm eq}=1.0$
and $T=0.1$.

%\section{Results}
\begin{figure}
\includegraphics[width=8.5cm, angle=0, clip]{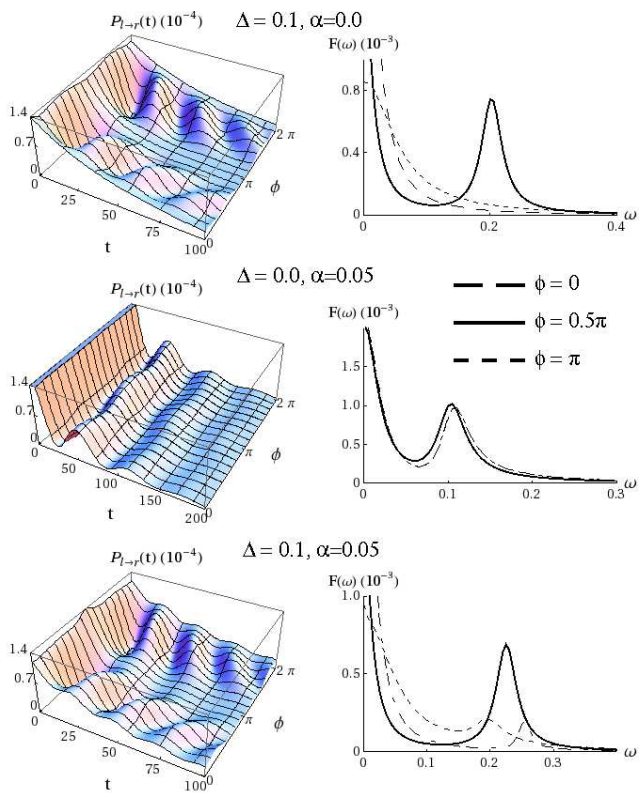}
\caption{
$P_{l\rightarrow r}(t)$ (left--panels) as functions of AB-phase $\phi$ and time $t$;
The corresponding Fourier transformation $F(\omega)$ (right--panels) at
$\phi=0$ (dash), $\pi/2$ (solid), and $\pi$ (dot), respectively.
The upper, middle and bottom panels are for three representing
sets of interdot-transfer rate $\Delta$ and orbital detuning $\alpha$.
Other parameters are $U=1.0, T=0.1, 2V=0.2$ and $\mu_{\rm eg}=1.0$
(in unit of $\epsilon_g$); see text for details.
\label{fig:r1}}
\end{figure}

  Figure \ref{fig:r1}
demonstrates the dependence of the waiting time distributions $P_{l \rightarrow r}(t)$
(left--panels) on the AB-phases $\phi$,
together with their Fourier
transforms $F(\omega)$ (right--panels)
exemplified at three representing values of
$\phi=0, \pi/2$, and $\pi$.
The Coulomb repulsion parameter $U=1.0$ and the bias $2V=0.2$
are common, while
the interdot transfer and
orbital energy split parameters are
$(\Delta,\alpha) = (0.1, 0)$ in the upper,
$(0,0.05)$ in the middle, and
$(0.1,0.05)$ in the bottom panels, respectively.
Clearly, the influences of $\Delta$ and $\alpha$
on the waiting time distribution
are {\it qualitatively}  distinct,
especially in the two limiting
regimes. While $P_{l \rightarrow r}(t)$ shows only
little dependence on $\phi$ in
the dot orbital--split case (the middle panel:
$\Delta= 0$ but $\alpha\neq 0$),
it is strikingly sensitive to
the AB-phase in the interdot--transfer case
(the upper panel: $\Delta\neq 0$ but $\alpha=0$).
In the latter case, the characteristic
oscillation is maximized at $\phi=\pi/2$,
but disappears at $\phi=0$ and $\phi=\pi$.
These observations can be largely understood as follows.

 The dot orbital--split ($\Delta= 0$ but $\alpha\neq 0$)
case resembles the transport through double slits.
The resulting interference \cite{Wel08195315}
 persists and is insensitive
to the AB phase, due to
the fact that $\phi_l=-\phi_r$ in each orbital channel
largely cancels out the AB--phase effect.
This accounts for the basic feature observed in the
middle panels of Fig.\,\ref{fig:r1}.

In interdot--transfer
case (upper panels: $\Delta\neq 0$ but $\alpha=0$),
the aforementioned double-slit feature
is destroyed. The interdot-transfer allows electrons
to switch between the two pathways provided by the DQD.
Thus, different phases can be accumulated as
the electron transfer through the coupled
DQD and the aforementioned phase symmetry is broken
along some of the possible transfer trajectories.
As a result, the total accumulated phase depends
on the value of $\phi$.  It leads to
a pure decay of $P_{l \rightarrow r}(t)$
at the AB phase $\phi=0$ or $\pi$, where
$e^{i\phi}=e^{-i\phi}$.
However, at other values of $\phi$,
it leads to an effective phase difference
between the eigen-levels which are subject to an induced energy gap
of $2 \Delta$, responsible for
the AB--phase activated
oscillations observed in
the upper panels of Fig.\,\ref{fig:r1}.

In the intermediate regime shown in the bottom panels of Fig.\,\ref{fig:r1},
oscillations can be observed for all $\phi$;
however, the Fourier transform
reveals a frequency shift when AB-phase is tuned by the magnetic field.
At $\phi=\pi/2$, the observed frequency corresponds to the DQD
eigenenergy gap $(2 \sqrt{\Delta^2+\alpha^2}=0.224)$, while at
$\phi=0$ or $\pi$, it is blue or red shifted, respectively.
The amplification of the oscillation at $\phi=\pi/2$ is characteristic for inter-dot
transfer and allows to distinguish it from orbital detuning.
The latter causes only small oscillations at $\phi=0$ or
$\pi$.

\begin{figure}
\includegraphics[width=8.5cm, angle=0, clip]{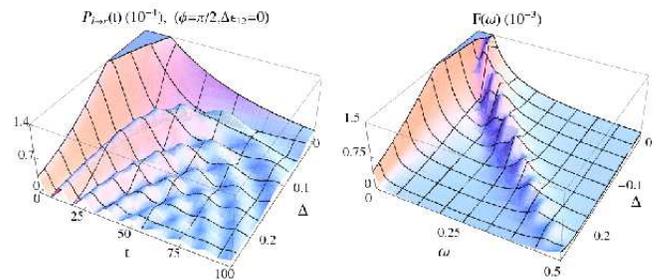}
\caption{
$P_{l\rightarrow r}(t)$ as function of inter-dot transfer
rate $\Delta$ and time $t$ and corresponding
Fourier transformation $F(\omega)$. The AB-phase is $\phi=\pi/2$
and the orbital detuning $\alpha=0$. Other parameters are
same as Fig.\ \ref{fig:r1}.
\label{fig:r2}}
\end{figure}

Figure \ref{fig:r2}
examines further the influence of $\Delta$
on $P_{l \rightarrow r}(t)$  (left)
and its spectrum $F(\omega)$ (right), with
$\alpha=0$ and $\phi=\pi/2$, where oscillations due to
AB phase--activated interferences between the eigen-levels
are at maximum.
Note that the interference would remain dark at $\phi=0$ in this case;
as can be seen the upper panels of Fig.\,\ref{fig:r1}.
The amplitude of $P_{l\rightarrow r}(t)$ oscillation decreases
with $\Delta$, which corresponds to a decreased average
current through the DQD.

%\fi

%The challenge in determining $P_{l \rightarrow r}(t, t_0)$
%analytically is to calculate the propagator $S_{t,t_{0}}$ in

\begin{figure}
\includegraphics[width=8.0cm, angle=0, clip]{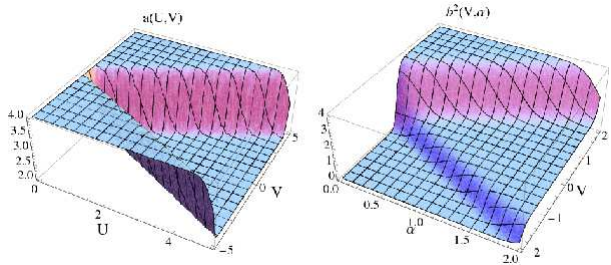}
\caption{The damping parameter $a(U,V)$ and
 the pre-exponential amplitude parameter $b^2(V,\alpha)$,
as their function dependence,
 for the oscillation term $P_{osc}(t)$ [\Eq{osc1}].
Inter-dot electron transfer is absent $\Delta=0$.
\label{fig:r0}}
\end{figure}

To analyze other coherent operation conditions,
%(besides $\phi=\pi/2$),
let us focus on the orbital--detuning
($\Delta=0$ and $\alpha\neq 0$) case
 where $\tilde{H}_S$ is diagonal.
We also neglect the Liouville--space off--diagonal elements in
$\Pi_{l}+\Pi_{r}$, which have a relatively small
influence in the weak coupling
regime. As results, the propagator $S_{t,t_{0}}$
in determining $P_{l \rightarrow r}(t, t_0)$
[\Eq{lr1}] becomes diagonal,
and the analytical solution is achievable.
Moreover, the waiting time distribution is separable into
$P_{l\rightarrow r}(t)=P_{osc}(t)+P_{decay}(t)$.
A detailed discussion of the decaying
terms $P_{decay}(t)$ that depends weakly on the AB--phase would
exceed the scope of this letter; but it has been provided for the
case of incoherent transport through single benzene molecules
\cite{Wel081137} which can be applied to QD-systems as well.
As the coherent operation conditions are concerned, we
focus only on the
oscillation term, which is independent of the AB--phase
for the orbital--split case (cf.\ the middle panels of Fig.\,\ref{fig:r1}).
\begin{equation}\label{osc1}
P_{osc}(t)= p_0\Gamma^2 b^2(V,\alpha) e^{-2a(U,V) \Gamma t}
\mathrm{cos}(2\alpha t),
\end{equation}
where $p_0$ is the initial vacuum state occupation number,
$a(U,V)=a(U,-V)=f(U+V)+ f(U-V)+2$, and
\begin{equation}
b(V,\alpha)=\frac{1+e^{2 \alpha \beta }+2 e^{(V+\alpha )
 \beta } }{e^{\beta(V+2 \alpha+1)/2}} f(V+\alpha)f(V-\alpha) .
\end{equation}
The left--panel of Fig.\ \ref{fig:r0} depicts the damping parameter
$a$ as function of $U$ and $V$.
It assumes the maximum value of 4
for small Coulomb coupling $U < V$ and is independent of $\alpha$.
Also the decay rate proportional to the system-electrode
coupling strength $\Gamma$. Thus a weak coupling is required
for the observability of interferences. This qualifies
statistical analysis of waiting time distributions as an indirect
method to study internal processes indirectly avoiding fast decoherence
in the system.

 The right--panel of Fig.\ \ref{fig:r0} depicts
the pre-exponent parameter
$b^2$ as function of $V$ and $\alpha$.
It reveals further the parameter regimes where
oscillations are observable.
One condition is that $V<\alpha$. Oscillations
are suppressed at negative bias larger than the DQD energy gap.
The amplitude is strongly increased when $V>\alpha$. However in this regime
the decay rate $2a \Gamma$ may reach its maximum
and prevent the observability of
coherence.
Apparently, the presence of strong Coulomb coupling,
as well as operating at small bias regime,
are favored for the observation of interference
effects by means of waiting time distributions.

In conclusion, a Markovian quantum master equation
in the Fock space was formulated and
employed to calculate the waiting time distribution
of consecutive electron transfers in AB interferometers.
Based on this we describe a novel statistical method
to determine quantum interferences,
inter-dot electron transfers, orbital detuning and the AB-phase.
Orbital detuning and inter-dot transfer induce oscillations
in the waiting time distribution in the presence
of interference. The two cases can be distinguished qualitatively
since the latter one is sensitive to the AB-phase.
The observability of oscillations requires
the presence of strong Coulomb interaction,
small bias and a weak electrode-system coupling.

The indirectness of the statistical detection avoids fast decoherence but
a large number of transfer events is necessary in order to extract information.
This might be advantageous for a qubit in operation with a continuous
readout. The method does not provide
information on a single operation but can determine whether
a large set of operations is carried out coherently.
Also other sources of decoherence like coupling to phonon bath
have to be minimized. The signature of interferences in
waiting time distribution can survive in the presence
 of a phonon-bath \cite{Bra08477}.
The proposed scheme can be realized utilizing presently available technology.
For that purpose two DQD in series
junctions which act as detectors by their coupling to their
respective quantum point contacts \cite{Fuj061634}
should be installed on both sides of a parallel DQD junction.
This setup consisting of six QDs avoids decoherence
inflicted by the charge state measurement of the quantum point contact.

Support from the RGC (604007 \& 604508) of Hong Kong (to YJY),
NSF (CHE-0745892/CBC-0533162) and
NIRT (EEC 0303389) of USA (to SM)
 is acknowledged.

%\bibliography{/home/yan/refs/bibrefs}
%\bibliographystyle{/home/yan/refs/aip}
%\bibliographystyle{aip}
%\bibliography{bibrefs}
%\bibliography{paper2008-3}
%\bibliography{bibrefs}

\end{document}